\RequirePackage{fix-cm}
\documentclass[smallextended]{svjour3}       
\smartqed  
\usepackage{float,subfigure}
\usepackage{amsmath,amssymb,bm}
\usepackage{color}
\usepackage{latexsym}
\usepackage{epstopdf}
\usepackage{multirow}
\usepackage{graphics,graphicx}
\DeclareGraphicsExtensions{.bmp,.jpg,.eps,.pdf,.png}
\usepackage{float,verbatim}
\usepackage{soul}
\usepackage{lineno,hyperref}
\usepackage[square,sort,comma]{natbib}

\usepackage[toc,page]{appendix}
\newcommand{\bx}{\bm{x}}

\usepackage{lineno}

\journalname{}
\begin{document}

\title{A History Matching Approach for Calibrating Hydrological Models}


\author{Natalia V. Bhattacharjee \and
        Pritam Ranjan \and
        Abhyuday Mandal \and
        Ernest W. Tollner
}


\institute{Natalia V. Bhattacharjee \at
              Institute for Health Metrics and Evaluation, University of Washington, Seattle, USA \\
              *Work was done at the Department of Statistics, University of Georgia, Athens, USA \and
            Pritam Ranjan \at
              OM\&QT, Indian Institute of Management Indore, MP, India \and
            Abhyuday Mandal \at
              Department of Statistics, University of Georgia, Athens, USA
              \email{amandal@stat.uga.edu} \and
            Ernest W. Tollner \at
              College of Engineering, University of Georgia, Athens, USA
}

\date{Received: date / Accepted: date}

\maketitle

\begin{abstract}

Calibration of hydrological time-series models is a challenging task since these models give a wide spectrum of output series and calibration procedures require significant amount of time. From a statistical standpoint, this model parameter estimation problem simplifies to finding an inverse solution of a computer model that generates pre-specified time-series output (i.e., realistic output series). In this paper, we propose a modified history matching approach for calibrating the time-series rainfall-runoff models with respect to the real data collected from the state of Georgia, USA. We present the methodology and illustrate the application of the algorithm by carrying a simulation study and the two case studies. Several goodness-of-fit statistics were calculated to assess the model performance. The results showed that the proposed history matching algorithm led to a significant improvement, of $30\%$ and $14\%$ (in terms of root mean squared error) and $26\%$ and $118\%$ (in terms of peak percent threshold statistics), for the two case-studies with Matlab-Simulink and SWAT models, respectively.

\end{abstract}

\keywords{History matching \and Contour estimation \and Gaussian process model \and Inverse problem \and Prediction \and Hydrology}

\section{Introduction}
\label{sec:Introduction}

Hydrological models are commonly used in environmental studies to estimate the water cycle elements in an area of interest. These models use basic principles of mass balance, energy conservation and other principles of physics. The input parameters of these models are often unknown and correspond to physical properties that are difficult to measure. Tuning/calibration of these parameters is required to obtain realistic outputs \citep{montanari2007}. This calibration problem is also referred to as the inverse problem in computer experiments literature. This research deals with obtaining the set of input parameters of a computer model that corresponds to a pre-specified target response, which is the observed field data in our application.

In this paper, we focus on calibrating two time-series valued hydrological models that simulate rainfall-runoff dynamics. The input parameters of these models are high dimensional, and the outputs can be very sensitive to small changes in the inputs. Realistic computer models can also be computationally and/or financially expensive, which prohibits numerous evaluation of the simulator. As a result, the calibration of these time-series models is a challenging problem, and an efficient approach to find the inverse solution is extremely important. Several researchers have attempted to solve the inverse problem for hydrological models using different methods via both manual and automated approaches, such as, the Genetic Algorithms, Maximum Likelihood Estimator, Markov Chain Monte Carlo, and Shuffled Complex Evolution \citep{duan1992, franchini1997, boyle2000, montanari2007, chu2010, tigkas2015}.

In an unrelated endeavour, \cite{ranjan2016} and \cite{zhang2018inverse} proposed a sequential design strategies for estimating the inverse solution, and \citet{vernon2010} proposed an iterative approach called history matching (HM) for calibrating a galaxy formation model called GALFORM. HM algorithm intelligently eliminates the implausible points from the input (or parameter) space and returns a set of plausible candidates for the inverse solution. However, there are a few aspects of the HM algorithm by \cite{vernon2010} that differ from our objective. First, the end result of the HM algorithm may be an empty set if there does not exist a plausible inverse solution, and second, the HM algorithm requires a large number of simulator runs which is undesirable in several applications like ours, where the simulator is expensive to evaluate.

We propose a modification in the HM algorithm which allows us to find the inverse solution in fewer simulator runs, and gives us a perfect match if possible, otherwise, the best approximation instead of returning an empty set of inverse solutions. We carry out a simulation study and two case studies of rainfall-runoff models to apply the proposed algorithm in solving this inverse mapping problem. To the best of our knowledge, the HM algorithms have not been applied yet for calibration of hydrological models with time series response.

The case studies refer to the calibration of two rainfall-runoff computer simulators for two target data sets collected at different locations in the state of Georgia, USA, which contains forty to fifty windrow composting systems. The management of the composting pad is crucial since the pad runoff is highly regulated and researchers have tried to estimate runoff in order to provide guidance for retention pond design \citep{kalaba2007, wilson2004}. The first case study focusses on the calibration of Matlab-Simulink compartmental dynamic model that estimates the amount of runoff from the windrow composting pad \citep{duncan2013instantaneous}. We wish to calibrate this model with respect to the composting pad data from the Bioconversion center, University of Georgia, Athens. The second case study considers the calibration of Soil and Water Assessment Tool (SWAT) model, a complex hydrological model that simulates runoff from watershed areas based on climate variables, soil types, elevation and land use data \citep{arnold1994}. We use the Middle Oconee River data for calibrating this model. SWAT is an internationally accepted simulator and used in modeling of the rainfall-runoff processes across various watersheds and river basins to address climate changes, water quality, land use and water resources management practices \citep{krysanova2015, dile2013, jayakrishnan2005, srinivasan2005}.

The rest of the manuscript is organized as follows. Section~\ref{sec:methods} presents the methodology for the proposed history matching algorithm for solving the inverse problems. Section~\ref{sec:simulation-study} presents a simulation study. The implementation of the proposed strategy is shown for the two case studies in Sect.~\ref{sec:application}. Section~\ref{sec:Discussion} concludes the article with a summary and important remarks.

\section{Methodology}
\label{sec:methods}

Let $g(\bx) := \{g(\bx,t_i), i=1,2,...,L\}$ denote the time-series valued simulator response for a given input $\bx \in [0,1]^d$ (scaled to an unit hypercube for convenience). Then the objective of the inverse problem is to find the $\bx$ (or set of $\bx$'s) that generate the desired (pre-specified) output $g_0 := \{g_0(t_i), i=1,2,...,L\}$ (say). For many complex phenomena, the realistic computer models are also computationally and/or financially expensive to run. As a result, standard mathematical techniques and algorithms cannot be used for solving the inverse problems. \cite{ranjan2008} proposed a sequential design approach for efficiently finding the inverse problem for scalar-valued simulators. However, for this research, the complexity due to time-series response makes the problem more challenging. Section~\ref{sec:HM-algorithm} briefly reviews the history matching (HM) algorithm proposed by \citet{vernon2010}, and then we discuss the proposed modifications to the HM algorithm in Sect.~\ref{sec:HM-algorithm-modified}.

\subsection{History Matching Algorithm}
\label{sec:HM-algorithm}

The history matching algorithm proposed by \citet{vernon2010} begins by discretizing the time-series response on $T_k$ time points, say, at $t_1^*, t_2^*, ..., t_{T_k}^*$, such that $T_k$ is much smaller than $L$. These $T_k$ time points are chosen in such a way that they capture the defining features of the target response. Then, the HM method finds a common set of plausible solutions to these $T_k$ inverse problems for scalar-valued simulators, and declares it as a solution to the general inverse problem. Mathematically, the HM algorithm finds $\bx \in [0,1]^d$  such that $g(\bx,t_j^*) = g_0(t_j^*)$ for all $j=1,2,...,T_k$.

Assuming that the computer model is expensive, the inverse solution must be estimated using the minimal number of model runs. A common practice in computer experiments literature is to build up the methodologies using a flexible statistical surrogate trained on carefully chosen model runs. \cite{vernon2010} used the most popular surrogate, Gaussian process (GP) model. For simplicity, let us assume that $y(\bx_i) = g(\bx_i, t_j^*)$. Then, the $n$ training points, $(\bx_i,y(\bx_i)), i=1,2,...,n$, are modelled as $ y(\bx_i) = \mu + Z(\bx_i)$, where $\mu$ is the mean and $\{Z(\bx), \bx \in [0,1]^d\}$ is a GP, denoted by $Z(\bx)\sim GP(0, \sigma^2R)$. This implies that $E(Z(\bx))=0$ and the spatial covariance structure defined as $Cov(Z(\bx_i), Z(\bx_j)) = \Sigma_{ij} = \sigma^2 R(\theta; \bx_i,\bx_j)$.  [Notation: We use bold $\bx_i$ to denote a $d$-dimensional point in $[0,1]^d$ and un-bold $x_{ik}$ to denote the $k$-th coordinate of $\bx_i$.]

For any given input $\bx^*$ in the design space, the fitted GP surrogate gives the predicted simulator reponse as,
\begin{equation}
	\label{eqn-gp-estimation}
	\hat{y}(\bx^*) = \mu + \mathbf{r}(x^*)^T\mathbf{R}^{-1}(\mathbf{y}-\mu\mathbf{1}_n),
\end{equation}
where $\mathbf{r}(\bx^*) = [\text{corr}(z(\bx^*), z(\bx_1)), \text{corr}(z(\bx^*), z(\bx_2)),...,\text{corr}(z(\bx^*), z(\bx_n))]^T$, $\mathbf{1}_n$ is a vector of ones of length $n$, $\mathbf{R}$ is the $n\times n$ correlation matrix for $(Z(\bx_1), ..., Z(\bx_n))$, $\mathbf{y}$ is the response vector $(y(\bx_1),...,y(\bx_n))$, and the associated uncertainty estimate is,
\begin{equation}
	\label{eqn-gp-estimated-variance}
	s^2(\bx^*) = \sigma^2\left(1 - \mathbf{r}(\bx^*)^T\mathbf{R}^{-1}\mathbf{r}(\bx^*)\right).
\end{equation}
In practice, the parameters $\mu, \sigma^2$ and $\theta$ in Equations~(\ref{eqn-gp-estimation}) and (\ref{eqn-gp-estimated-variance}) are replaced by their estimates (see \cite{vernon2010} for details). We used the \texttt{R} package \texttt{GPfit} \citep{gpfit} for obtaining $\hat{y}(\bx^*)$ and $s^2(\bx^*)$ for any arbitrary $\bx^*$ and a given training data.

The driving force behind the HM algorithm is the implausibility function
\begin{equation}\label{eqn-imp-function}
 I_{(j)}(\bx) = \frac{|\hat{g}(\bx,t_j^*)-g_0(t_j^*)|}{s_{t_j}(\bx)} ,
\end{equation}
where $\hat{g}(\bx,t_j^*)$ is the predicted response in Equation~(\ref{eqn-gp-estimation}), and $s_{t_j}(\bx)$ is the associated uncertainty estimate in Equation~(\ref{eqn-gp-estimated-variance}). The main idea is to label the design points implausible if $I_{max}(\bx) > c$, where
$$ I_{max}(\bx) = \max\{I_{(1)}(\bx), I_{(2)}(\bx), ..., I_{(T_k)}(\bx)\},$$
and $c$ is a pre-determined cutoff (e.g., $c=3$ as per $3\sigma$ rule of thumb). \citet{vernon2010} further proposed an iterative approach to refine the plausible subset of points from the input space. However, the algorithm is designed to find the set of all plausible inverse solutions and not only the perfect solution. For the Galaxy formation model (GALFORM) application with input dimension $d=17$, \citet{vernon2010} used a large training set to start with ($n_1 = 1000$) and ended up with $N=2011$ points after four iterations.

\subsection{Modified History Matching Algorithm}
\label{sec:HM-algorithm-modified}

We propose a few modifications in the history matching algorithm described above. We aim to find only the best possible approximation of the inverse solution instead of the entire plausible set, and prefer to use a reasonably small space-filling design instead of a large design in $[0,1]^d$ for building the initial surrogate. The optimal choice for the size of design, $n_1$, is discussed in Section~3.2. The key steps of the proposed modified HM algorithm are summarized as follows:

\begin{enumerate}
\item[1.] \emph{Choose a discretization-point-set (DPS), $t_1^*, t_2^*, ..., t_{T_k}^*$.}
\item[2.] \emph{Set $i=1$. Assume $D_0 = \phi$ (empty set).}
\item[3.] \emph{Choose a training set, $D_1 = \{\bx_1,\bx_2,...,\bx_{n_1}\} \subset [0,1]^d$, using a space-filling design, and evaluate the simulator $g(\bx)$ over $D_1$.}
\item[4.] \emph{Fit $T_k$ scalar-response GP-based surrogate to $g(\bx,t_j^*)$ over the training set $D = D_i \cup D_{i-1}$.} We used the \texttt{R} package \texttt{GPfit} for surrogate fitting.

\item[5.] \emph{Evaluate the implausibility criteria $I_{(j)}(\bx)$ for $j=1,2,...,T_k$ over a randomly  generated test set $\chi_i$ of size $M$ (via a space-filling design) in $[0,1]^d$ and combine them via
$$ I_{max}(\bx) = \max\{I_{(1)}(\bx), I_{(2)}(\bx), ..., I_{(T_k)}(\bx)\},$$
for screening the plausible set of points $D_{i+1} = \{\bx \in \chi_i\ : \ I_{max}(\bx) \le c\}$.}

\item[6.] \emph{Stop if $D_{i+1} = \phi$, otherwise, set $i=i+1$, evaluate the simulator on $D_i$ and go to Step~4.}
\end{enumerate}

Instead of using the entire $D_{i+1}$ from Step~5 to Step~6, one can use a space-filling design to find a representative subset of $D_{i+1}$ and then augment it in Step~4 for the next iteration. This will further reduce the total computer model evaluation in solving the inverse problem. Since we assume that the target response is a realization of the simulator output, one can find the best possible approximation of the inverse solutions by minimizing the discrepancy $\delta(\bx)=\|g(\bx) - g_0\|$, where $\|\cdot\|$ is the Euclidean distance or $L_2$ norm. Assuming $N$ is the total number of points at the end of the proposed HM algorithm, the desired inverse solution is given by 
$$\hat{\bx}_{opt} =  \underset{1\le i \le N}{\textrm{argmin}} \ \|g(\bx_i) - g_0\|.$$
Instead of minimizing $\delta(\bx)$ over the training set, one can develop an extraction technique using the final fitted surrogate and/or the DPS. 

In summary, we need to identify the following elements to implement the proposed history matching algorithm:

\begin{itemize}
\item[(a)] \emph{a computer model} ($g(\cdot)$) that takes a $d$-dimensional input vector and returns a time-series output,

\item[(b)] \emph{input parameters} ($\bx$) that need to be calibrated,

\item[(c)] \emph{a target response} ($g_0$) for calibrating the computer model, and 

\item[(d)] \emph{algorithmic parameters}: $n_1, c, T_k, (t_1^*,...,t_{T_k}^*)$ and $M$.

\end{itemize}

Next, we present a simulation study for a comprehensive understanding of the calibration problem and investigate different aspects of the proposed algorithm. Two real-life case studies are presented in Sect.~\ref{sec:application}.

\section{Simulation Study}
\label{sec:simulation-study}

\begin{figure}[h!]\centering
        \includegraphics[width=3.75in]{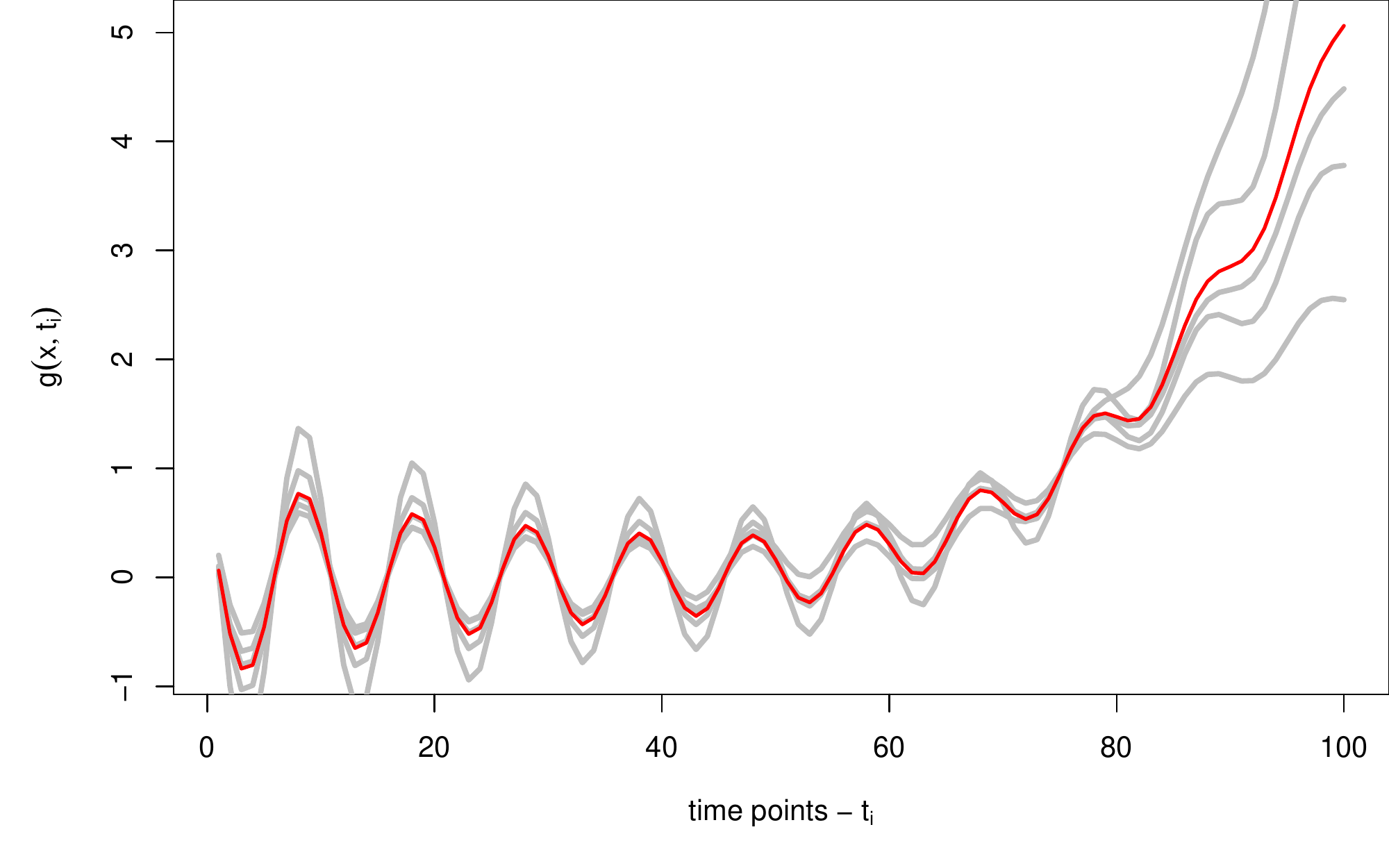}
        \caption{The illustrative example: a few model outputs (dashed curves) and the target response (solid curve).}
        \label{fig:illus_example}
\end{figure}

The objective of this simulation study is to discuss the implementation details of the proposed algorithm, and investigate the sensitivity of the algorithmic parameters on the performance efficiency. We consider a simple test function as a computer simulator with two calibration parameters. Specifically, the inputs are $\bx=(x_1,x_2) \in [0,1]^2$, which return the following time-series output:
\begin{equation}\label{eq:test-function}
g(\bx,t_i) = \frac{\sin(10 \pi t_i)}{(2x_1+1)t_i} + |t_i-1|^{(4x_2+2)},
\end{equation}
where  $t_i=0.5,0.52,0.54, ..., 2.50$ (equidistant time points of length $L=101$ in $[0.5, 2.5]$). We further assume that the true value of the calibration parameter is $\bx_0=(0.5,0.5)$, which generates the target response $g_0$ in the inverse problem context. Figure~\ref{fig:illus_example} presents the model outputs for a few random input combinations (gray curves) and the target response series (red curve).

Our objective is to find $\bx\in [0,1]^2$ such that $g(\bx)\approx g_0$. We now apply the proposed HM algorithm for solving the inverse problem.

\subsection{Application of the Proposed Algorithm}
\label{sec:simulation-method-comparison}

The implementation procedure stats with choosing the algorithmic parameters. Since the computer simulator, as shown in Figure~\ref{fig:illus_example}, appears to be quite simple and $d=2$, we wish to start with $n_1=10$ points for fitting the initial surrogate (note that the choice of $n_1$ is somewhat arbitrary at this point). The cutoff for selecting the plausible points is chosen as $c=3$, which is guided by the $3\sigma$ rule of thumb for normal distributions. We randomly selected $T_k=2$ and then used $L/3$ and $2L/3$ for discretizing the response, i.e., $DPS = (33, 67)$, since $L=101$. Finally, we used a randomly chosen large dense sets of size $M=5000$ for thoroughly searching the follow-up points in the subsequent iterations. That is, the algorithmic parameters are: $n_1 = 10$, $c=3$, $T_k=2$, $DPS = (33, 67)$ and $M=5000$.  

\begin{figure}[h!]\centering
\includegraphics[width=4.5in]{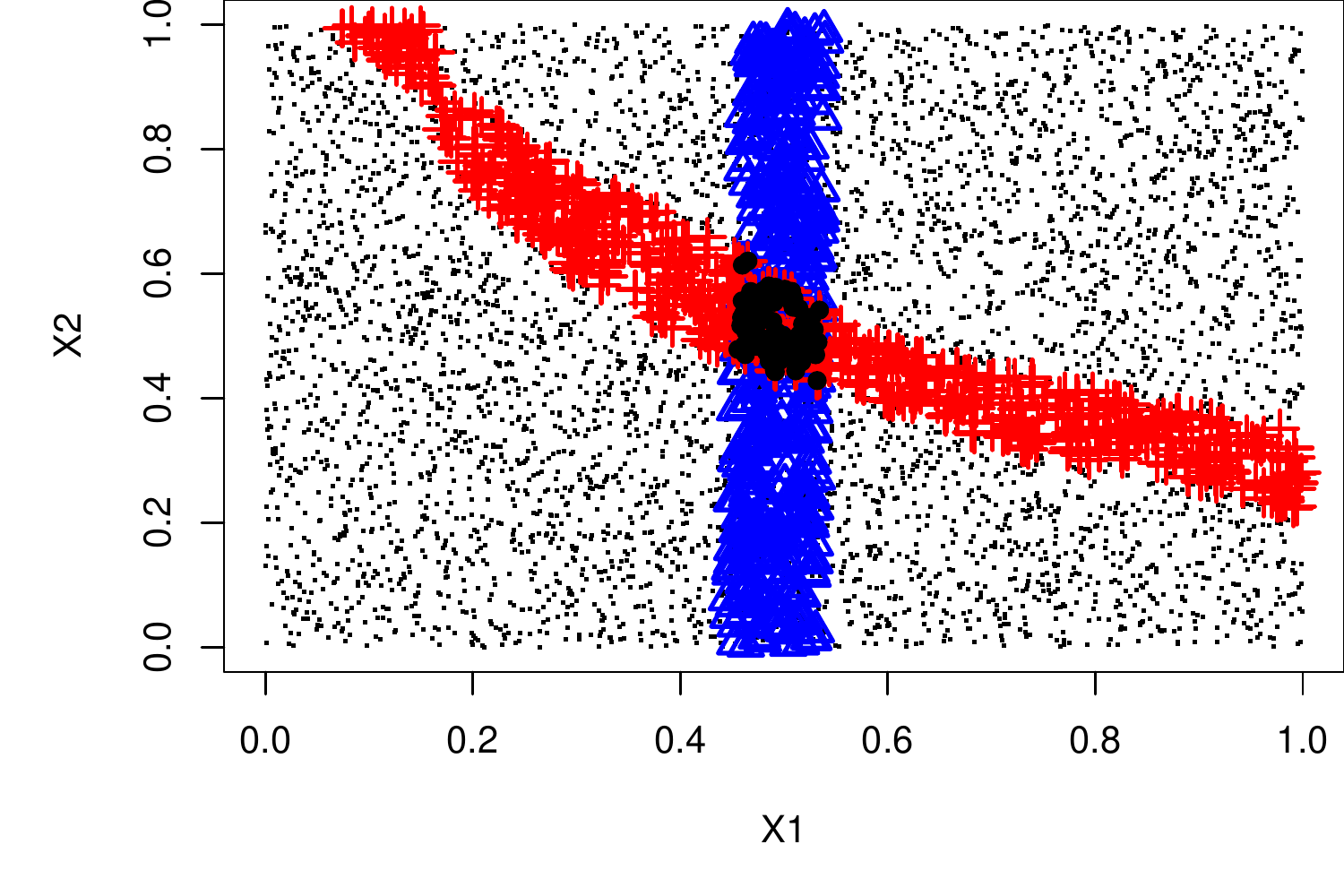}
\caption{The illustrative example: selection of the training points according to the implausibility function with cutoff $c=3$ at the discretization-point-set $DPS=(33, 67)$ in the first iteration of the modified HM algorithm.} \label{fig:illus_hm_wave-a}
\end{figure}

Figure~\ref{fig:illus_hm_wave-a} provides the selection of points in the first iteration, where the points in (blue) triangle and (red) plus correspond to $I_{(j)}(\bx)\le 3$ for $t_1^*=33$ and $t_2^*=67$ respectively, and the (black) solid circle represents $D_2=\{I_{max}(x)\le 3\}$. The iterative procedure gives $|D_2|=69$. 

\begin{figure}[h!]\centering
\includegraphics[width=4.5in]{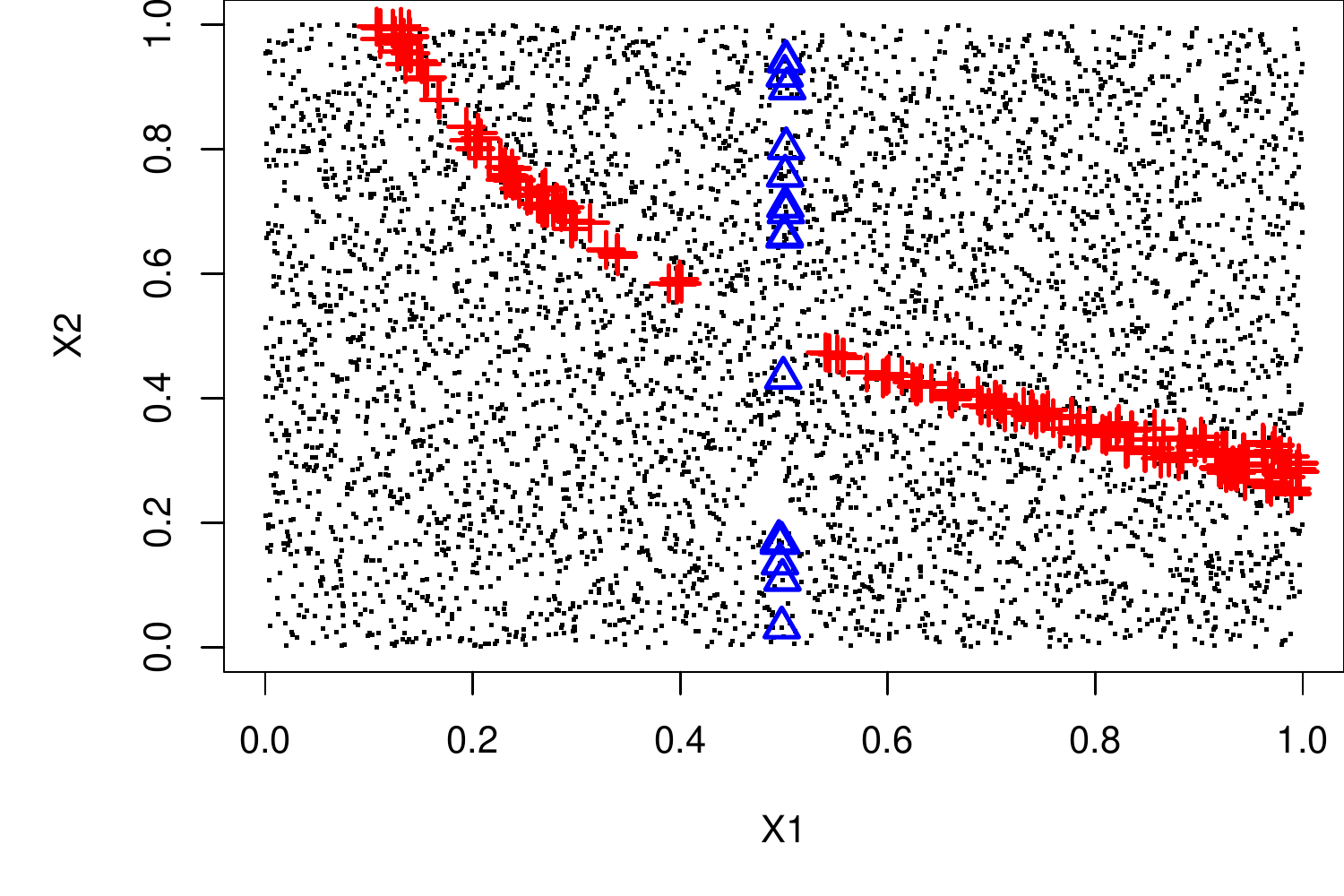}
\caption{The illustrative example: selection of the training points according to the implausibility function  with cutoff $c=3$ at the discretization-point-set $DPS=(33, 67)$ in the second iteration of the modified HM algorithm.} \label{fig:illus_hm_wave-b}
\end{figure}

Subsequently, the augmented training set is of size $79$. Now, for the second iteration, Figure~\ref{fig:illus_hm_wave-b} shows the implausibility value of the candidate points. It turns out that $D_3$ is an empty set, i.e., there are no black solid dots in this figure. This happens because individually $\{x\ :\  I_{(j)}(\bx)\le 3\}$ are non-empty for both $j=1,2$, but $I_{max}(\bx) \not\le 3$. Thus, the iterative procedure terminates.

\begin{figure}[h!]\centering
        \includegraphics[width=3.5in]{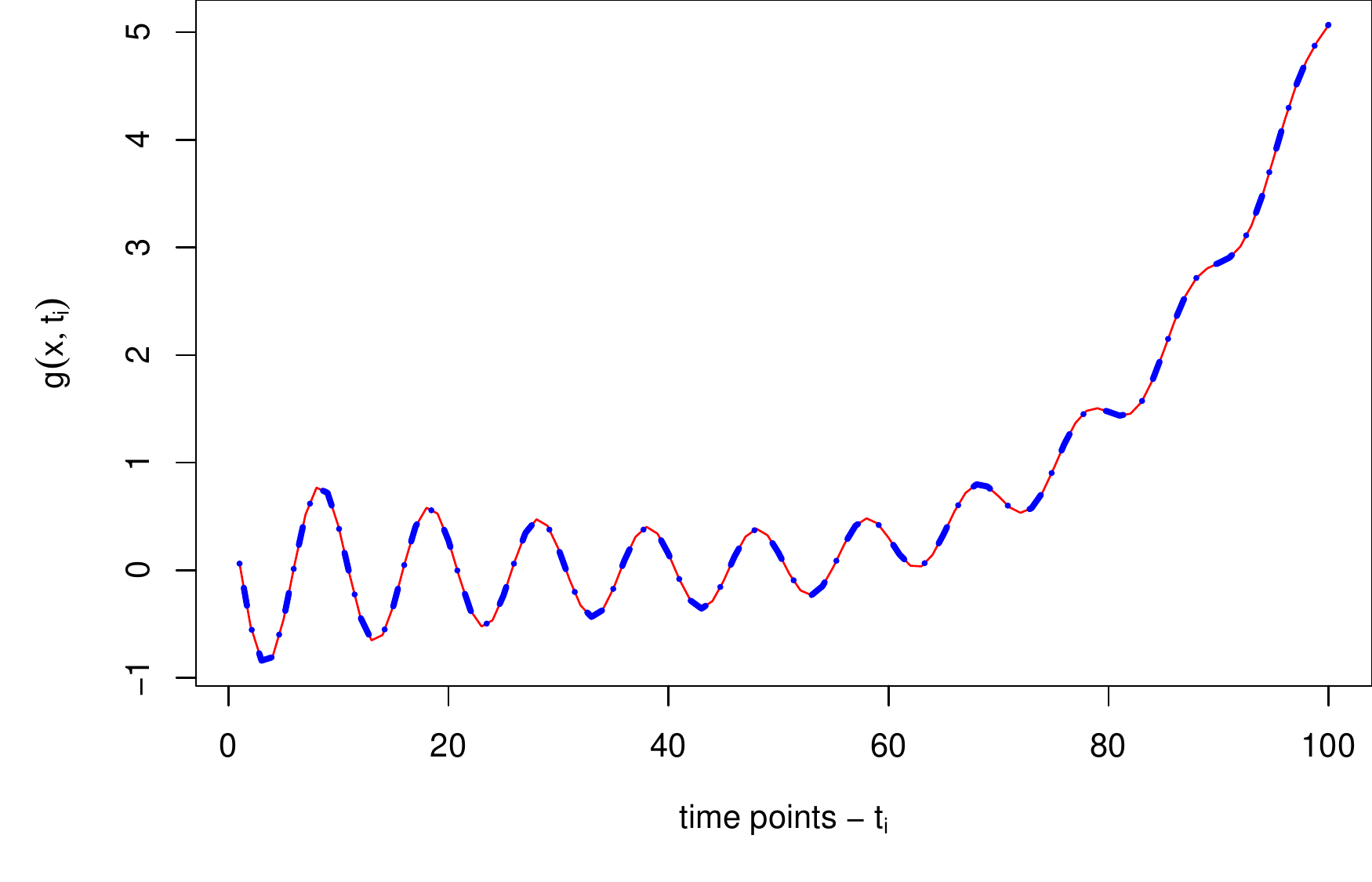}
        \caption{The illustrative example: the simulator output at the estimated inverse solution $\hat{\bx}_{opt}$ (dashed blue curve) and the target response (solid red curve).}
        \label{fig:illus_best}
\end{figure}

As a result, the final training set is of size $N=79$, and the minimized $\log[\delta(\bx_i)]$ over the training set is $-4.2290$, with the estimated inverse solution $\hat{\bx}_{opt}=(0.4992, 0.5007)$. It turns out that the simulator output at $\hat{\bx}_{opt}$ is very similar to the target response (see Figure~\ref{fig:illus_best}).

\subsection{Sensitivity of Algorithmic Parameters}
\label{sec:simulation-method-efficiency}

We now investigate the sensitivity of the algorithmic parameters, $n_1, c, T_k$ and $M$, with respect to the accuracy of the estimated inverse solution measured by $\log[\delta(\hat{\bx}_{opt})]$, which is the minimized value of $\delta(\bx)$ over the augmented training data at the end of the proposed HM algorithm. That is, the lower the value of $\log[\delta(\hat{\bx}_{opt})]$, the better the parameter combination is. We randomly regenerated the initial training sets, test sets and the DPS for each combination of $n_1=(5,10,20)$, $c=(1,2,3)$, $T_k=(2,4,8)$ and $M=(500, 2000, 5000)$, and ran the modified HM algorithm. The results are averaged over 100 random realizations for each combination of $n_1, c, T_k$ and $M$.

\begin{figure}[!h]\centering
\includegraphics[width=4.5in]{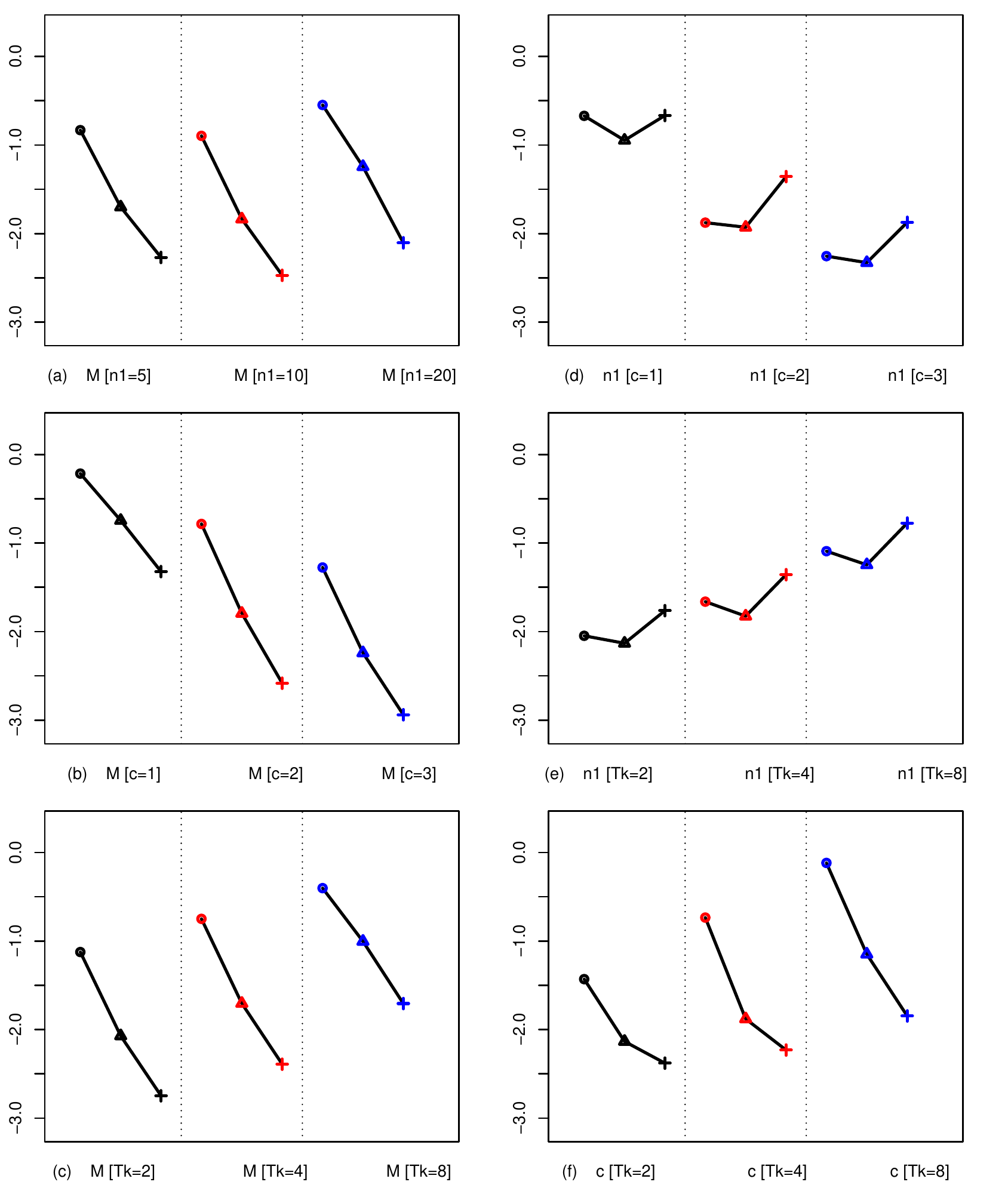}
\caption{The illustrative example: marginal distribution of the median of $\log[\delta(\hat{\bx}_{opt})]$ over 100 simulations for different two-factor combinations of $n_1, c, T_k$ and $M$.}\label{fig:hm_sim_illus_bart_gp_comparison}
\end{figure}

Figure~\ref{fig:hm_sim_illus_bart_gp_comparison} presents the marginal distribution of the median of $\log[\delta(\hat{\bx}_{opt})]$ over 100 simulations for all possible two-factor combinations of $n_1, c, T_k$ and $M$. Here, each panel has three sub-panels. For Panel (a), the left most sub-panel corresponds to $n_1=5$ and the three dots there correspond to $M=500$ (solid circle), $M=2000$ (solid triangle), and $M=5000$ (plus), respectively. Similarly, the middle sub-panel shows the different values of $\log[\delta(\hat{\bx}_{opt})]$ for the same three different values of $M$ and a fixed value of $n_1$(=10). The line segments in other panels and sub-panels can be explained similarly. 

From Figure~\ref{fig:hm_sim_illus_bart_gp_comparison} we can draw some inference regarding the sensitivity and preference for the algorithmic parameters. For example, Panels (a), (b) and (c) show that as the value of $M$ increases, from 500 to 5000, the value of $\log[\delta(\hat{\bx}_{opt})]$ decreases monotonically. Naturally, here $M=5000$ is the best choice. Although it may not be obvious from Panel (a), Panels (d) and (e) clearly demonstrate that $n_1 = 10$ give better results for this example, since in all of these cases, the value of $\log[\delta(\hat{\bx}_{opt})]$ for $n_1=10$ is smaller than that of $n_1=5$ or $20$. Similarly, Panels (b) and (d) support the choice of $c=3$, and the same conclusion can be drawn from Panel (f), since each of the three lines of this panel has the lowest value of $\log[\delta(\hat{\bx}_{opt})]$ at $c=3$. Finally, Panels (c), (e) and (f), all clearly indicate that $T_k=2$ gives the lower value of $\log[\delta(\hat{\bx}_{opt})]$ than that for $4$ and $8$.

Together, these six panels of Figure~\ref{fig:hm_sim_illus_bart_gp_comparison} lead to some intuitive conclusions, such as the higher the value of $M$ or $c$, the better the performance of the proposed HM algorithm. However, some other conclusions are not that intuitive, and these simulations shed more light on the optimal choice of the algorithmic parameters. For example, it turns out that a higher number of dicretized points ($T_k$) may not necessarily yield a better performance of the HM algorithm. Finally, if the size of the initial design is either too small or too large, the HM algorithm will not be very efficient. It is important to note that the inferences drawn here are based only on this small simulation study for a simple test function based simulator, and the optimal choices for the algorithmic parameters will have to be carefully chosen for another application.

\begin{figure}[ht]\centering
\includegraphics[width=4in]{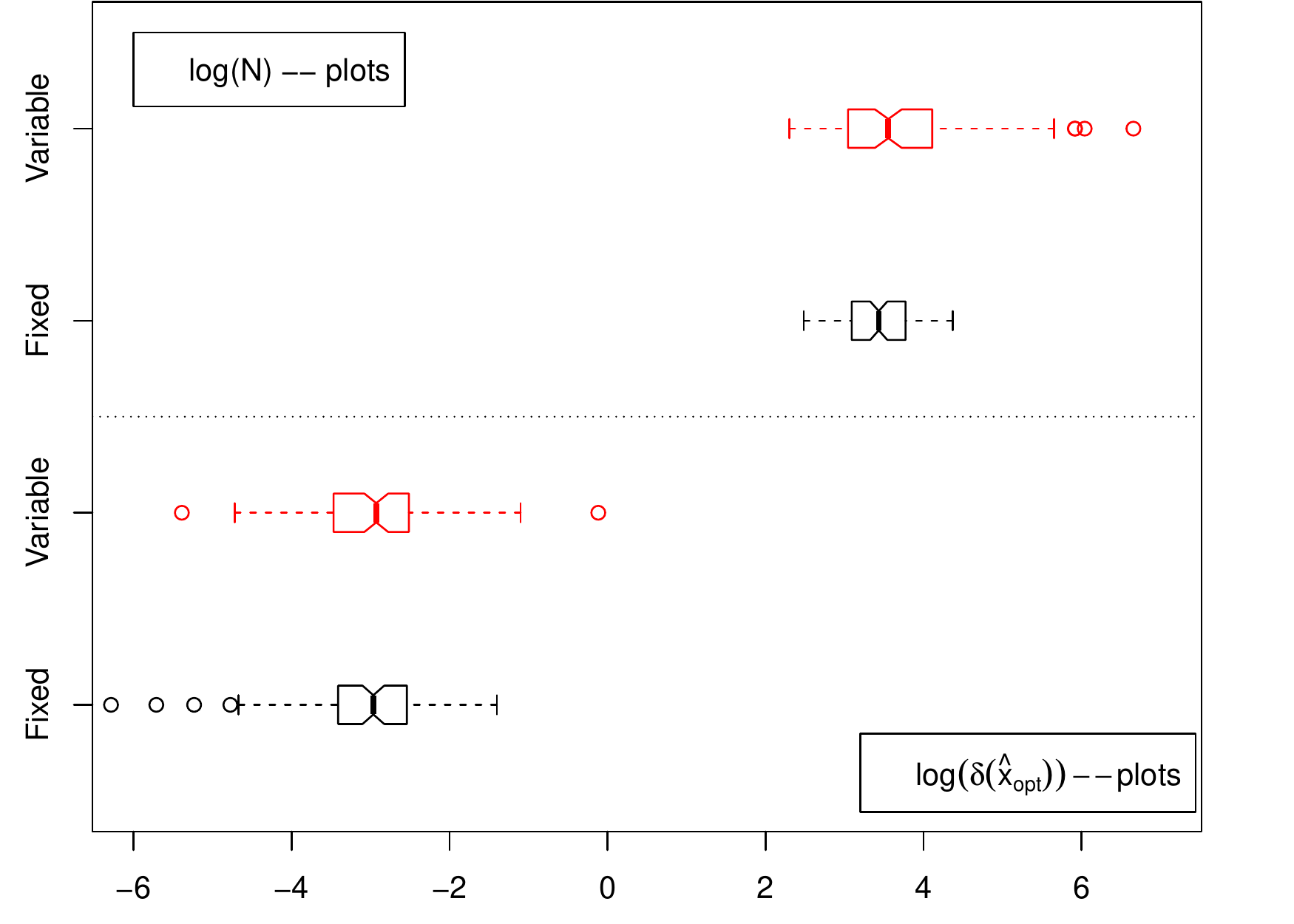}
\caption{The illustrative example: Sensitivity of selecting DPS measured with respect to the total run-size and optimized $\log[\delta(\hat{\bx}_{opt})]$.}\label{fig:illus_hm_Tp}
\end{figure}

Since the size of the discretization-point-set (value of $T_k$) plays a crucial role in the performance of HM algorithm, the actual location of the discretization points (i.e., DPS) may also affect the performance of the proposed algorithm. Figure~\ref{fig:illus_hm_Tp} presents the performance comparison of the proposed algorithm over 100 simulations. Here, we fix $n_1=10, c=3$ and $T_k=2$, and randomly generate training data and implement the algorithm under two scenarios: \textit{Fixed} $-$ DPS=$(33,67)$, and \textit{Variable} $-$ randomly generate DPS of size $T_k$ using some space-filling criterion. The top panel of Figure~\ref{fig:illus_hm_Tp} presents $\log(N)$ distribution and the bottom panel displays $\log[\delta(\hat{\bx}_{opt})]$  distribution over 100 simulations for both fixed and variable scenario.

It is clear from the top panel of Figure~\ref{fig:illus_hm_Tp} that the choice of DPS fixed at $(33, 67)$ is clearly better than many other alternatives in terms of the total number of computer model evaluations. The bottom panel shows that both scenarios \textit{Fixed} and \textit{Variable} give comparable accuracy of the final inverse solution, which is expected as the termination of the algorithm depends on the accuracy of the predictor near the target response, as captured by the implausibility function in Equation~(\ref{eqn-imp-function}). In summary, a good choice of the DPS may be helpful in efficiently finding the inverse solution.

\textbf{Remark 1:} For real-life applications, it is certainly infeasible to experiment with different choices of DPS to find the optimal one. One would have to carefully choose DPS to ensure that the important features are captured. The objective of the above simulation study is to demonstrate that the choice of DPS is important for estimating $\hat{\bx}_{opt}$ with the fewest number of computer model simulator runs. 

\textbf{Remark 2:} A reasonable choice of $n_1$ is also a non-trivial problem. It varies with the end objective, complexity of the underlying simulator response process and the input dimension. In an attempt to answer this question, \citet{loeppky2006} suggests a rule of thumb of 10 points per input dimension to be enough for getting a good overall idea of the underlying process (i.e., $n_1=10d$, where $d$ is the input dimension).  However, our objective is to estimate the inverse solution only and not to explore the entire input space with same accuracy. Thus the choice of $n_1=10d$ is not necessarily optimal in our case. In a sequential design framework for estimating pre-specified features of interest, e.g., global minimum or the inverse solution, \cite{ranjan2008} recommends using $n_1 \in [N/3, N/2]$ for building the initial surrgoate.

\section{Case studies}
\label{sec:application}

This section illustrates the implementation of the proposed history matching approach for the calibration of two hydrological models. The first case study deals with Matlab-Simulink model which simulates runoff from windrow compost pad over a period of time. The second case study refers to estimating the inverse solution of a well-known reservoir model called Soil and Water Assessment Tool (SWAT). 

\subsection{Case Study $1$: Matlab-Simulink Model}
\label{sec:matlab-application}

\citet{duncan2013instantaneous} investigated the rainfall-runoff relationship for the windrow composting pad, and developed a compartmental model for estimating the amount of runoff from the composting pad (represented as a change in pond volume). It quantifies the surface runoff, infiltration and lateral seepage using differential equations developed for each section of the compost pad. Additionally, the model takes several factors as inputs, for instance, length, width, slope of compost pad, area covered by compost windrows, depth of surface/sub-surface, depression/embankment depths, initial surface/sub-surface water content, and model coefficients of the saturated hydraulic conductivity of the gravel media ($K_{sat1}$) and the saturated hydraulic conductivity of the supporting soil below the media ($K_{sat2}$). As per \citet{duncan2013instantaneous}, the following four inputs/parameters are the most influential: depth of surface, depth of sub-surface and two coefficients of the saturated hydraulic conductivity ($K_{sat1}$ and $K_{sat2}$). See \citet{duncan2013instantaneous} for more details on data collection, characteristics of composting pad and the Matlab-Simulink model. 

\begin{figure}[h!]\centering
 \includegraphics[width=5in]{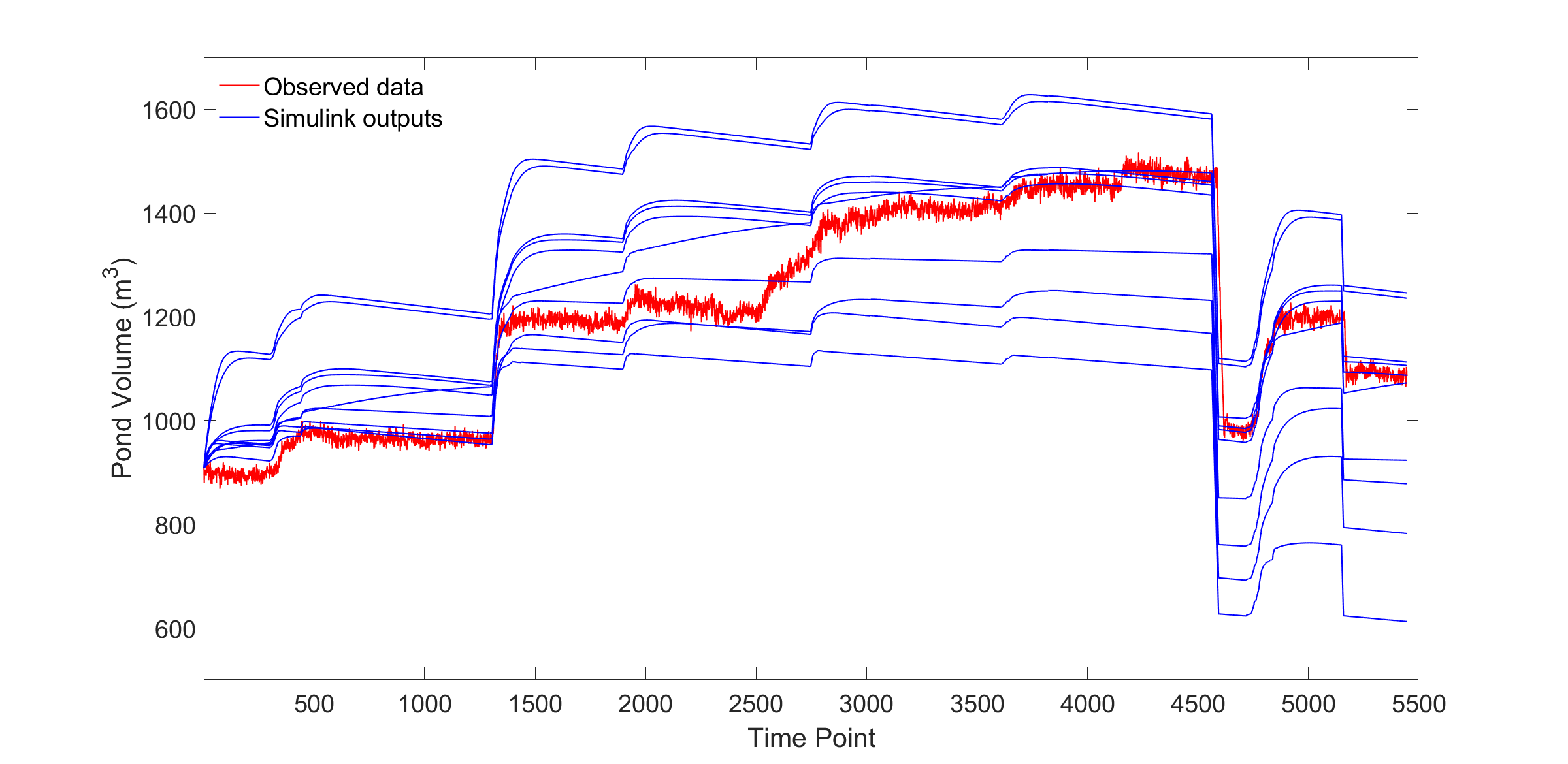}
\caption{Field data ($g_0(t_i)$) from Bioconversion center, UGA (represented by the red curve) and the Matlab-Simulink model outputs $g(\bx, t_i)$ (represented by the blue lines) for $i= 1, 2, ..., 5445$ at randomly generated $\bx$ (depth of surface, depth of sub-surface and two coefficients of the saturated hydraulic conductivity $K_{sat1}$ and $K_{sat2}$ ). Time period is December 23, 2010 - January 30, 2011.}\label{Fig:msm_illustration}
\end{figure}

For calibration, we used the runoff data ($g_0$) collected at Bioconversion center, University of Georgia, Athens, USA, as the target response. The raw runoff data (collected on a $10$-minute interval during 11:50AM, December $23$, 2010 to 11:50PM, January $30$, $2011$ over $T=5445$ time points) are represented by the noisy (red) curve in Figure~\ref{Fig:msm_illustration}. This figure shows a few random computer model responses superimposed with the field data. 

The descriptive statistics of the field data required to compute the runoff are summarized in Table~\ref{tab:matlab-var-summary-stat}.
\begin{table}[h!]
\begin{center} \caption{Summary Statistics of the field data (collected at Bioconversion center, University of Georgia, Athens, USA) required for the Matlab-Simulink Model case study.}\label{tab:matlab-var-summary-stat}
\begin{tabular}{|l|cccccc|}
\hline
\multicolumn{1}{|c|}{Variable (units)} & \multicolumn{6}{|c|}{Summary} \\
& {Min} &  {Median} &  {Mode} & {Mean} & {Std} & \multicolumn{1}{c|}{Max}\\
\hline
Rainfall (cm)      & 0  &   0   &  0  &   0.002  &  0.011  &  0.345 \\
Pond Volume ($m^3$)     & 867.50 & 1203.00  & 1192.40 &   1207.20  &  191.40 &   1515.90  \\
\hline
\end{tabular}
\end{center}
\end{table}

The objective here is to find the best possible combinations of those four inputs / parameters: depth of surface, depth of sub-surface, $K_{sat1}$ and $K_{sat2}$, that can generate realistic runoff, i.e., similar to the one obtained from the field data. For convenience in the implementation of the algorithm, the inputs were scaled to $[0,1]^4$. We start the proposed HM algorithm implementation by choosing $n_1 = 40$ points using  a maximin Latin hypercube design \citep{johnson1990}, and evaluate the simulator on these design points. By carefully examining the nature of the field data, five time points ($T_k=5$) given by $\{135, 554, 1243, 3232, 4500\}$  were selected from the runoff series (of length $L=5445$) to discretize the time-series responses. Furthermore, we used the test set of size $M=5000$ and $c=3$ for computing the implausibility values and finding the training points for the next iteration. The full implementation required $N=461$ simulator runs to converge.

\begin{figure}[h!]\centering
  \includegraphics[width=5in]{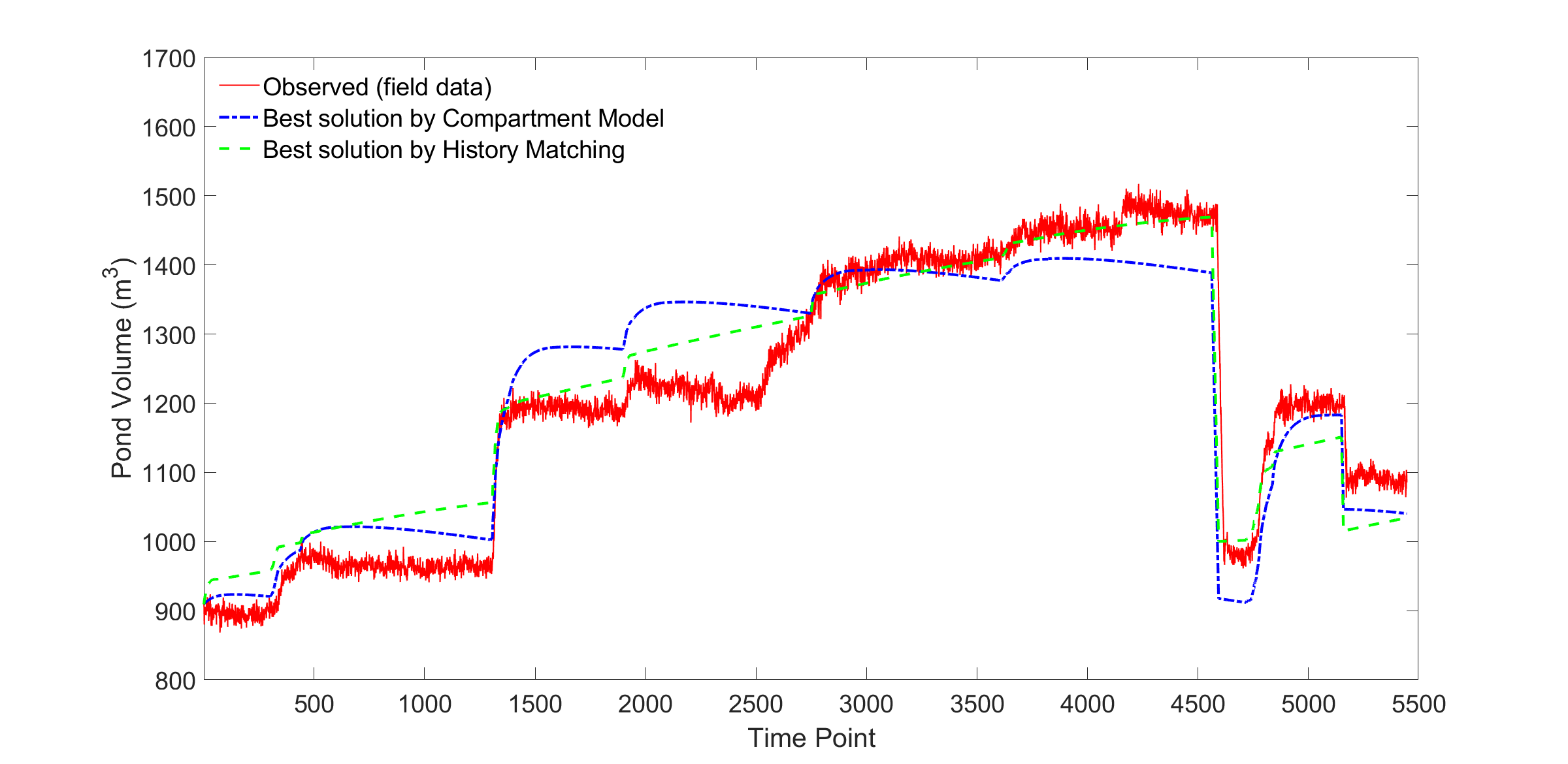}
  \caption{Calibration Results for the Matlab-Simulink model. The solid red curve represents observed data, blue dash line represents best solution used in the previous study and green dash line corresponds to the best solution using the proposed HM algorithm. Time period is December 23, 2010 - January 30, 2011.}\label{Fig:simulink_results}
\end{figure}

The final inverse solution obtained via the proposed HM algorithm is presented in Figure~\ref{Fig:simulink_results}. For a benchmark comparison, we also present the best inverse solution found by \citet{duncan2013instantaneous}.

%


For accuracy comparison of different approaches, there are several goodness of fit measures that are more popular in hydrological applications as compared to $\log[\delta(\hat{\bx}_{opt})]$. We use four such popular measures in this article:
\begin{itemize}
\item Root mean squared error
$$RMSE  = \left(\frac{1}{L}\sum_{i=1}^{L}\left|g(\hat{\bx}_{opt},t_i)-g_0(t_i)\right|^2\right)^{1/2}.$$

\item Coefficient of determination $R^2$ of the simple linear regression (SLR) model, when the dependent variable is the target response and the independent variable is the estimated inverse solution, i.e., $R^2$ of the SLR model: 
$$g_0(t_i) = g(\hat{\bx}_{opt},t_i) + \varepsilon_i, i=1,2,...,L,$$
with the assumption of i.i.d. errors $\varepsilon_i$.

\item Nash-Sutcliffe Efficiency \citep{nse1970}
$$NSE = 1 - \frac{\sum_{i=1}^L[g(\hat{\bx}_{opt},t_i)-g_0(t_i)]^2}{\sum_{i=1}^L[g_0(t_i)-\bar{g}_0]^2}.$$

\item Peak percent threshold statistics \citep{lohani2014}: $PPTS_{(l, u)}$ is the trimmed mean of 
$$|\xi_{t_i}| = \frac{|g_0(t_i)- g(\hat{\bx}_{opt},t_i)|}{|g(\hat{\bx}_{opt},t_i)|}$$ 
after eliminating the two tail percentiles, $l\%$ and $u\%$, values of $|\xi_{t_i}|$.

\end{itemize}

Table~\ref{tab:matlab-performance} summarizes the values of these four goodness of fit measures for the calibration of Matlab-Simulink Model using the proposed HM algorithm and the state-of-the-art  Compartmental model \citep{duncan2013instantaneous}. For PPTS values we compute measures under two scenarios: no-trimming, and 5\% trimming each at the two tails. Note than $R^2$ and NSE should be maximized, whereas the other two statistics, RMSE and PPTS, should be minimized.

\begin{table}[h!]\centering \caption{Goodness of fit comparisons of the proposed HM algorithm and Compartmental model \citep{duncan2013instantaneous}  for calibrating the Matlab-Simulink Model.}\label{tab:matlab-performance}
\begin{tabular}{|lc|ccccc|}
\hline
&Matlab-Simulink& RMSE & $R^2$ & NSE & PPTS$_{(5,95)}$ & PPTS$_{(1,100)}$\\
\hline
\multirow{2}{*}{} & Compartment       & 71.91 & 0.86 & 0.86  & 4.70 & 4.75 \\
                  & History Matching  & 55.58 & 0.93 & 0.92  & 3.71 & 3.77 \\
\hline
\end{tabular}
\end{table}

As per Table~\ref{tab:matlab-performance}, the proposed HM algorithm outperforms the earlier approach by \citet{duncan2013instantaneous} with respect to all three goodness of fit measures, and in particular by a significant $(71.91-55.58)/55.58\times 100 \approx 30\%$ margin according to RMSE, and $26\%$ margin as per $PPTS_{(1, 100)}$.

\subsection{Case Study $2$: SWAT Model}
\label{sec:swat-application}

SWAT model has been widely used for modeling the rainfall-runoff processes across various watersheds and river basins to address climate changes, water quality, land use and water resources management practices \citep{arnold1994, jayakrishnan2005, srinivasan2005, dile2013,  krysanova2015}. This hydrological model takes several inputs, for example, curve number ($CN$), groundwater delay ($GW_{delay}$), available water capacity ($AWC$), baseflow factor ($\alpha_{BF}$), Manning's coefficient ($\nu$), etc. Based on experts' advise and preliminary variable screening analysis using Sequential Uncertainty Fitting (SUFI2) toolkit, we identified the following five parameters for the calibration exercise: $\nu$, effective hydraulic conductivity in the channel ($K$), $GW_{delay}$, groundwater ``revap" coefficient ($GW_{revap}$) and $AWC$.  More details on SUFI2 can be found in \citet{abbaspour2004,abbaspour2007}.

The target response was retrieved from the historical monthly data of streamflow from the US Geological Survey (USGS) water data website for the Middle Oconee River, Georgia, during the period January 2001 to December 2009 (gauge number $02217500$). We obtained ASTER digital elevation model (DEM) values at $30$m resolution from USGS EarthExplorer platform and Global Climate Data in SWAT format from Texas $A\&M$ University website (https://globalweather.tamu.edu/). We used a warm-up period of two years (January $2001$ to December $2002$) and a calibration period of seven years (January $2003$ to December $2009$). For the stream flow records used in SWAT model, the descriptive statistics are listed in Table~\ref{tab:swat-var-summary-stat}.
\begin{table}[H]
\begin{center} \caption{Summary Statistics of the field data (stream flow records observed at $L=84$ time points for the Middle Oconee River, Georgia) used in the SWAT Model calibration.}\label{tab:swat-var-summary-stat}
\begin{tabular}{|l|cccccc|}
\hline
\multicolumn{1}{|c|}{Variable (units)} & \multicolumn{6}{|c|}{Summary} \\
& {Min} &  {Median} &  {Mode} & {Mean} & {Std} & \multicolumn{1}{c|}{Max}\\
\hline
Streamflow ($m^3/s$)    &  0.024  &   0.277   &  0.320  &   0.349  &  0.280  &  1.206 \\
\hline
\end{tabular}
\end{center}
\end{table}

\begin{figure}[h!]\centering
  \includegraphics[width=5.1in]{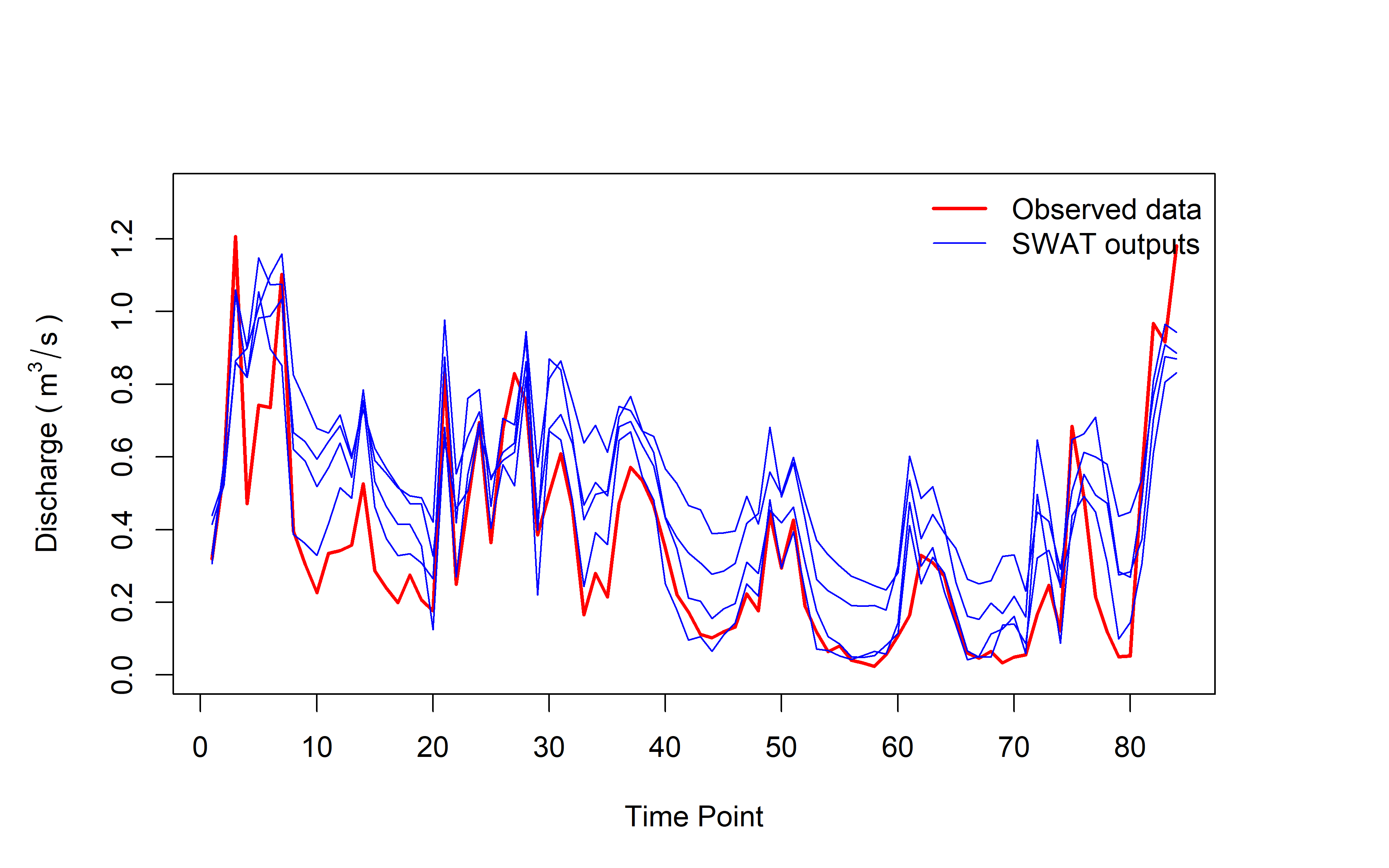}
  \caption{Middle Oconee river discharge data (USGS gauge number 02217500), $g_0(t_i)$
(red curve), and SWAT model discharge outputs $g(\bx, t_i)$ (blue curves) at random inputs $\bx$ (Manning's coefficient, effective hydraulic conductivity, groundwater delay, groundwater ``revap" coefficient and available water capacity). Time period is January 2003 - December 2009.}
\label{Fig:swat_illustration}
\end{figure}

Figure~\ref{Fig:swat_illustration} shows a few SWAT model runs (in blue $-$ obtained by randomly varying the calibration inputs) and the field data (in red).

\begin{figure}[h!]\centering
  \includegraphics[width=5.1in]{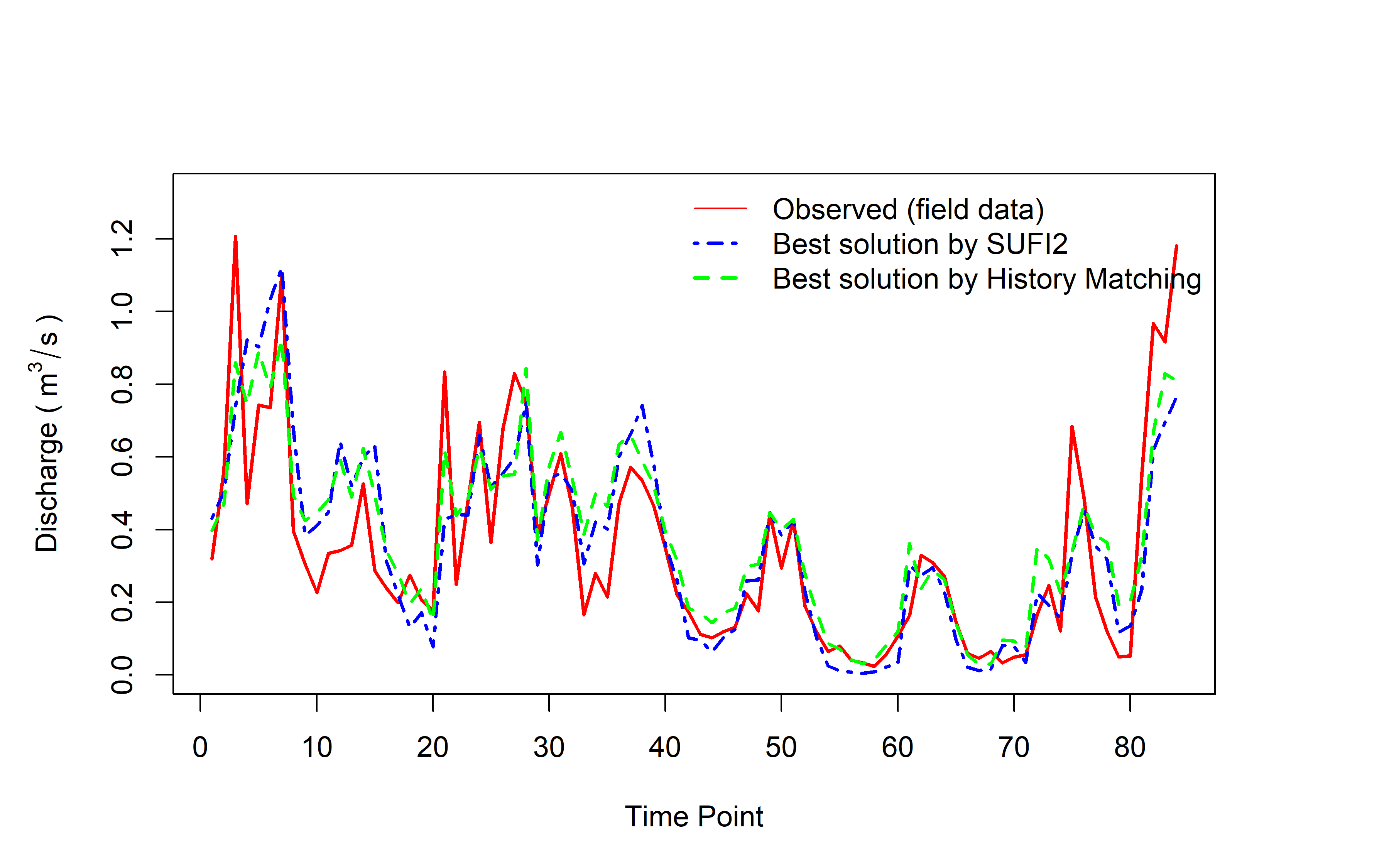}
  \caption{SWAT model calibration: The solid red curve represents the observed data, blue dashed line represents best solution using SUFI2, and the green dashed line corresponds to the best solution using the HM algorithm. Time period is January 2003 - December 2009.}\label{Fig:swat_results}
\end{figure}

Following the steps of the proposed HM algorithm (Sect.~\ref{sec:HM-algorithm-modified}), we rescaled the inputs to $[0,1]^5$, assigned $n_1 = 50$ for training the initial surrogate, and carefully identified four time instances $t_j^*$ at: $10, 37, 63, 79$ for discretizing the output series. The DPS contains two dips and two peaks. Here also we used test sets of size $M=5000$ and the cutoff for implausibility function to be $c=3$. Ultimately, the algorithm required $N=398$ model runs to converge. Figure~\ref{Fig:swat_results} presents the estimated inverse solution (dashed greed) along with the target response (solid red). For reference comparison, the best solution obtained by SUFI2 (dashed blue) has also been overlayed in Figure~\ref{Fig:swat_results}.

Table~\ref{tab:simulink-performance} presents a more detailed comparison of the two approaches measured with respect to RMSE, $R^2$, NSE and PPTS. Recall that $R^2$ and NSE have to be maximized and RMSE and PPTS have to be minimized.

\begin{table}[h!]\centering \caption{Accuracy comparisons of the proposed HM algorithm over the state-of-the-art Sequential Uncertainty Fitting (SUFI2) toolkit for the calibration of SWAT model.}\label{tab:simulink-performance}
\begin{tabular}{|c|ccccc|}
\hline
SWAT model & RMSE & $R^2$ & NSE & PPTS$_{(5,95)}$ & PPTS$_{(1,100)}$\\
\hline
 SUFI2             & 0.16 & 0.68 & 0.67  & 52.02 & 65.80\\
 History Matching  & 0.14 & 0.77 & 0.75  & 29.67 & 30.20\\
\hline
\end{tabular}
\end{table}

Similar to the previous case study, the proposed HM algorithm exhibits superior performance in terms of all four goodness of fit measures. In particular, the proposed approach demonstrates $(0.16-0.14)/0.14 \times 100 \approx 14\%$ improvement as per the RMSE criterion, and an amazing $118\%$ improvement with respect to $PPTS_{(1, 100)}$ measure.

\section{Discussion}
\label{sec:Discussion}

In this study, we applied the proposed modified history matching (HM) algorithm for solving an inverse problem (i.e. calibration problem) for a test function based computer model and two real-life hydrological models. The proposed algorithm demonstrated very good performance in all scenarios. In the first case study (Matlab-Simulink model), the HM algorithm demonstrated approximately 30\% better (as per RMSE) performance than the state-of-the-art compartment model calibration results. For the second case study, we observed that the HM algorithm resulted in approximately 14\% more accurate (as per RMSE) inverse solution as compared to the one obtained from SUFI2. Thus, we believe that the proposed HM algorithm can be fruitful for solving calibration problems in hydrological time-series models.

Based on our empirical findings via a simulation study, we infer that the choice of \emph{algorithmic parameters} gives a trade-off between large training-set and accuracy of the inverse solution. Due to the stochastic nature of the HM algorithm, a multi-start approach of the proposed HM algorithm may lead to improved accuracy, and subsampling of $D_i$ in Step~5 may lead to more economical sampling strategy, however one must analyze the tradeoff between the accuracy gain and the additional cost of simulator evaluation for the application at hand. The choice of discretization-point-set is subjective and a key to the success of this algorithm. In practice, one should examine the target response carefully, and choose the points in such a way that they capture the overall variation and important features reasonably well.

Note that the proposed HM approach will find the closest possible approximation in case the simulator turns out to be stochastic and cannot generate the exact same desired output $g_0$. Although, it is methodologically straightforward to generalize the proposed technique that can adjust for some systematic discrepancies, a bias correction step would require synchronised data on the simulator and actual field trials for multiple input combinations.


\section*{Acknowledgement}
The authors would like to thank the Editor, the Associate Editor and two reviewers for their thorough and helpful reviews. Ranjan's research was partially supported by the Extra Mural Research Fund (EMR/2016/003332/MS) from the Science and Engineering Research Board, Department of Science and Technology, Government of India. Mandal and Tollner's research was partially supported by 104B State Water Resources Research Institute Program, USA Grant G16AP00047. We would like to thank NASA DEVELOP National Program's node at the Center for Geospatial Research, UGA for providing resources on SWAT modeling.


\bibliographystyle{spbasic}      
\bibliography{bibfile}   



%
\end{document}